\documentclass{aa}  
\usepackage{graphicx}
\usepackage{txfonts}
\usepackage{hyperref}
\hypersetup{breaklinks=true,colorlinks=true,urlcolor=blue, linkcolor=blue, citecolor=blue}

\newcommand{\kms}{km\,s$^{-1}$}

\newcommand{\cms}{cm$^{-2}$}
\newcommand{\cmc}{cm$^{-3}$}

\begin{document}

   \title{ALMA chemical survey of disk-outflow sources in Taurus
     (ALMA-DOT)}

   \subtitle{III. The interplay between gas and dust in the protoplanetary disk of DG Tau}

   \author{L. Podio
          \inst{1}
          \and
A. Garufi
          \inst{1}
          \and
C. Codella
          \inst{1, 2}
          \and
D. Fedele
          \inst{1,3}
          \and
K. Rygl
          \inst{4}
          \and
C. Favre
          \inst{2}
          \and
F. Bacciotti           
          \inst{1}
          \and
E. Bianchi
          \inst{2}
          \and
C. Ceccarelli
          \inst{2}
          \and
S. Mercimek
          \inst{1, 5}
          \and
R. Teague
          \inst{6}
          \and
L. Testi          
          \inst{7}
          }

   \institute{INAF - Osservatorio Astrofisico di Arcetri, Largo E. Fermi 5, 50125 Firenze, Italy\\
              \email{lpodio@arcetri.astro.it}
         \and Univ. Grenoble Alpes, CNRS, IPAG, 38000 Grenoble, France
         \and INAF - Osservatorio Astrofisico di Torino, Via Osservatorio 20, I-10025 Pino Torinese, Italy
         \and INAF - Istituto di Radioastronomia \& Italian ALMA Regional Centre, via P. Gobetti 101, 40129 Bologna, Italy
         \and Universit{\`a} degli Studi di Firenze, Dipartimento di Fisica e Astronomia, Via G. Sansone 1, 50019 Sesto Fiorentino, Italy
         \and Center for Astrophysics | Harvard \& Smithsonian, 60 Garden Street, Cambridge, MA 02138, USA
         \and European Southern Observatory (ESO), Karl-Schwarzschild-Str. 2, 85748, Garching, Germany
             }

   \date{Received ...; accepted ...}

 
  \abstract
   {Planets form in protoplanetary disks and inherit their chemical composition. It is therefore crucial to understand the molecular content of protoplanetary disks in their gaseous and solid components.}
   {We aim to characterize the distribution and abundance of molecules in the protoplanetary disk of DG Tau and to compare them with its dust distribution.}
   {In the context of the ALMA chemical survey of Disk-Outflow sources in the Taurus star forming region (ALMA-DOT) we analyze ALMA observations of the nearby disk-outflow system around the T Tauri star DG Tau in H$_2$CO $3_{1,2}-2_{1,1}$, CS $5-4$, and CN $2-1$ emission  at an unprecedented resolution of $\sim 0\farcs15$, which means $\sim 18$ au at a distance of 121 pc.}
   {Both H$_2$CO and CS emission originate from a disk ring located at the edge of the 1.3~mm dust continuum. CS probes a disk region that is slightly further out with respect to H$_2$CO; their peaks in emission are found at $\sim 70$  and $\sim 60$ au, with an outer edge at $\sim 130$ and $\sim 120$ au, respectively. CN originates from an outermost and more extended disk/envelope region with a peak at $\sim 80$ au and extends out to $\sim 500$ au. H$_2$CO is dominated by disk emission, while CS  also probes two streams of material possibly accreting onto the disk with a peak in emission at the location where the stream connects to the disk. CN emission is barely detected and both the disk and the envelope could contribute to the emission. Assuming that all the lines are optically thin and emitted by the disk molecular layer in local thermodynamic equilibrium at temperatures of $20-100$ K, the ring- and disk-height-averaged column density of H$_2$CO is $2.4-8.6 \times 10^{13}$ \cms, that of CS is $\sim 1.7-2.5 \times 10^{13}$ cm$^{-2}$, while that of CN is $\sim 1.9-4.7 \times 10^{13}$ cm$^{-2}$. Unsharp masking reveals a ring of enhanced dust emission at $\sim 40$ au, which is located just outside the CO snowline ($\sim 30$ au).}
{Our finding that the CS and H$_2$CO emission is co-spatial in the disk suggests that the two molecules are chemically linked. Both H$_2$CO and CS may be formed in the gas phase from simple radicals and/or desorbed from grains. The observed rings of molecular emission at the edge of the 1.3~mm continuum may be due to dust opacity effects and/or continuum over-subtraction in the inner disk, as well as to increased UV penetration and/or temperature inversion at the edge of the millimeter(mm)-dust which would cause enhanced gas-phase formation and desorption of these molecules. CN emission originates only from outside the dusty disk, and is therefore even more strongly anti-correlated with the continuum, suggesting that this molecule is a good probe of UV irradiation. The H$_2$CO and CS emission originate from outside the ring of enhanced dust emission, which also coincides with a change in the linear polarization orientation at 0.87~mm. This suggests that outside the CO snowline there could be a change in the dust properties that manifests itself as an increase in the intensity (and change of polarization) of the continuum and of the molecular emission.    
}


\authorrunning{Podio et al.}

   \titlerunning{ALMA-DOT II. Molecules and dust in the disk of DG Tau}

   \maketitle

%

\section{Introduction}

With the discovery of more than 4000 exoplanets, two key goals of modern astrophysics are to understand how planets form and what chemical composition they inherit from their natal environment. 
A viable way to answer these questions is to study protoplanetary disks around young Sun-like stars. The outstanding images recently obtained by the ALMA millimeter array provide the first observational indication of ongoing planet formation in disks of less than 1 Myr old, through rings and gaps in their dust and gas distribution \citep[e.g., ][]{alma15,andrews18,devalon20,favre19,fedele18,garufi20b,sheehan17,sheehan18}. The chemical composition of the forming planets clearly depends on the spatial distribution and abundance of molecules in the disk at the time of their formation. The chemical characterisation of  disks of 0.1-1 Myr old is therefore crucial.

This field has  long been hindered by observational difficulties due to the small sizes of disks ($\sim100$ au) and to the low gas-phase abundance of molecules (peak abundances with respect to H$_2$ down to 10$^{-12}$, e.g., \citealt{walsh14}). However, the number of disks imaged at high angular resolution in CO isotopologs \citep{booth19c,fedele17,isella16,zhang20} as well as in molecules other than CO has been rapidly increasing with ALMA. In particular, ALMA allows the radial distribution of small molecules to be 
retrieved, such as for example hydrocarbons C$_2$H, c-C$_3$H$_2$ \citep[e.g., ][]{bergin16,bergner19b,kastner15,loomis20,qi13c}, nitriles CN, HCN, HC$_3$N \citep{booth19b,hilyblant17,huang17,oberg15,vanterwisga19}, H$_2$CO \citep{carney17,carney19,garufi20b,guzman18a,kastner18,loomis15,oberg17,podio19,pegues20,qi13a,vanthoff20}, S-bearing molecules CS, H$_2$S, and H$_2$CS \citep[e.g., ][]{codella20,garufi20b,legal19,phuong18,teague18,loomis20,vanthoff20}, and molecular ions N$_2$H$^+$, DCO$^+$, H$^{13}$CO$^+$ \citep[e.g., ][]{booth19b,carney18,favre19,mathews13,oberg15b,qi13c}.  
The detection of complex organic molecules (COMs) is more difficult, and only a few COMs have been detected in non-bursting protoplanetary disks, such as CH$_3$CN, CH$_3$OH, and HCOOH \citep{bergner18,favre18,oberg15,podio20a,walsh16}. Other COMs, such as CH$_3$CHO and CH$_3$OCHO, have been detected in the disk of the FU Ori  outbursting  star V883 Ori \citep{lee-je19,vanthoff18}. In order to further our understanding in this area, we crucially need to enlarge the census of molecular distribution in disks, and to target younger disks because planet formation may occur earlier than previously thought. 
However, these disks are still partially embedded in their envelope and may be associated with molecular outflows. Therefore, an unprecedented combination of angular resolution and sensitivity is required to detect the faint emission from the disk and disentangle it from the other emitting components (e.g., the envelope and the outflow). To this aim we initiated the ALMA chemical survey of Disk-Outflow sources in the Taurus star forming region (ALMA-DOT program) \citep{codella20,garufi20b,podio19,podio20a}, which target Class I or early Class II disks associated with outflows in simple diatomic molecules (CO and CN), sulphur-bearing molecules (CS, SO, SO$_2$, H$_2$CS), as well as simple organics (H$_2$CO and CH$_3$OH) at $\sim20$ au resolution. 
The full sample, motivation, and overall results of ALMA-DOT are described in Garufi et al.\,in prep.

One of the sources targeted in the context of ALMA-DOT is the T Tauri star DG Tau ($d=121\pm2$ pc, \citealt{gaia16,gaia18}). DG Tau is surrounded by a compact and massive dusty disk imaged with CARMA \citep{isella10} and ALMA in polarimetric mode \citep{bacciotti18}.
Interferometric maps of CO and its isotopologs show that the envelope dominates over the molecular emission on large scales \citep{kitamura96a,schuster93} while disk emission is detected on scales $<2"$ \citep{gudel18,testi02,zhang20}.
The origin of the molecular emission detected with the IRAM 30m and {\it Herschel} \citep{fedele13,guilloteau13,podio12,podio13} is unclear because DG Tau is also associated with a residual envelope and a jet \citep{bacciotti00,eisloffel98}.
\citet{guilloteau13} suggest that the single-peaked profile of SO and H$_2$CO is due to envelope emission.
However, recent ALMA observations show that ALMA filters out extended molecular emission from the outflow or the envelope, thus isolating the compact emission from the disk, and show that H$_2$CO originates from a disk ring located at the edge of the dusty disk \citep{podio19}. 

In this paper, we present ALMA Cycle 4 observations of CS and CN molecules in the disk of DG Tau at an unprecedented resolution of $\sim 0\farcs15$, or $\sim 18$ au, we compare the distribution of these molecules with that of H$_2$CO analyzed by \citet{podio19} and with the dust distribution and substructures, and discuss the chemistry of these species.  


\begin{table*}
  \caption[]{\label{tab:lines} Properties of the observed lines and of the relative line cubes, integrated intensities, and estimated column densities.}
  \small{
  \begin{tabular}[h]{ccccccccc}
    \hline
    \hline
  Line & $\nu_{0}$$^{a}$ & E$_{\rm up}$$^{a}$ & S$_{ij} \mu^2$$^{a}$ & clean beam (PA) & $\Delta V$ & r.m.s.     & F$_{\rm int}$ & N$_{\rm X}$ \\
       & (MHz)           & (K)                &    (D$^2$)           &                 & (\kms)     & (mJy/beam) & (mJy \kms)    & ($10^{13}$ cm$^{-2}$)\\
    \hline 
o-H$_2$CO $3_{1,2}-2_{1,1}$          & 225697.775  & 33 & 43.5 & $0\farcs17 \times 0\farcs13$ ($-20\degr$)  & 0.16 & 1.7 & 210 & $1.8-5.5$\\
\\
CS $5-4$                             & 244935.557  & 35 & 19.1 & $0\farcs13 \times 0\farcs10$ ($-9.6\degr$)  & 0.6  & 0.6 & 352 & $1.7-2.5$ \\
\\
CN $2-1$, J=3/2-1/2, F=5/2-3/2$^{*}$ & 226659.5584 & 16 & 4.2  & $0\farcs14 \times 0\farcs12$ ($-7.6\degr$) & 0.16 & 1.5 & <75 & $<2-5$  \\
\multicolumn{1}{r}{J=3/2-1/2, F=1/2-1/2\,\,} & 226663.6928 & 16 & 1.2  &                                  &      &     &     &          \\
\\
CN $2-1$, J=5/2-3/2, F=5/2-3/2\,\,   & 226874.1908 &16 & 4.2   & $0\farcs14 \times 0\farcs12$ ($-7.8\degr$) & 0.16 & 1.5 & 173 & $1.9-4.7$ \\
\multicolumn{1}{r}{J=5/2-3/2, F=7/2-3/2$^{*}$} & 226874.7813 &16 & 6.7   &                                  &      &    &     &    \\
\multicolumn{1}{r}{J=5/2-3/2, F=3/2-1/2\,\,}   & 226875.8960 &16 & 2.5   &                                  &      &    &     &         \\
    \hline     
  \end{tabular}
  \\
$^{a}$ Molecular parameters from the CDMS database \citep{muller01}.\\
$^{*}$ The CN $2-1$ transition consists of 19 hyperfine structure components. The ALMA SPWs are centered on the brightest hyperfine components of CN $2-1$, indicated by an asterisk. Because of the line broadening due to disk kinematics, each of the CN 2-1, J=3/2-1/2, F=5/2-3/2, and CN 2-1, J=5/2-3/2, F=7/2-3/2 lines is blended with the adjacent hyperfine components reported in the table.  The integrated line intensity refer to the sum of the blended components.
}
\end{table*}



 

 \begin{figure*}
   \centering
   \includegraphics[width=\textwidth]{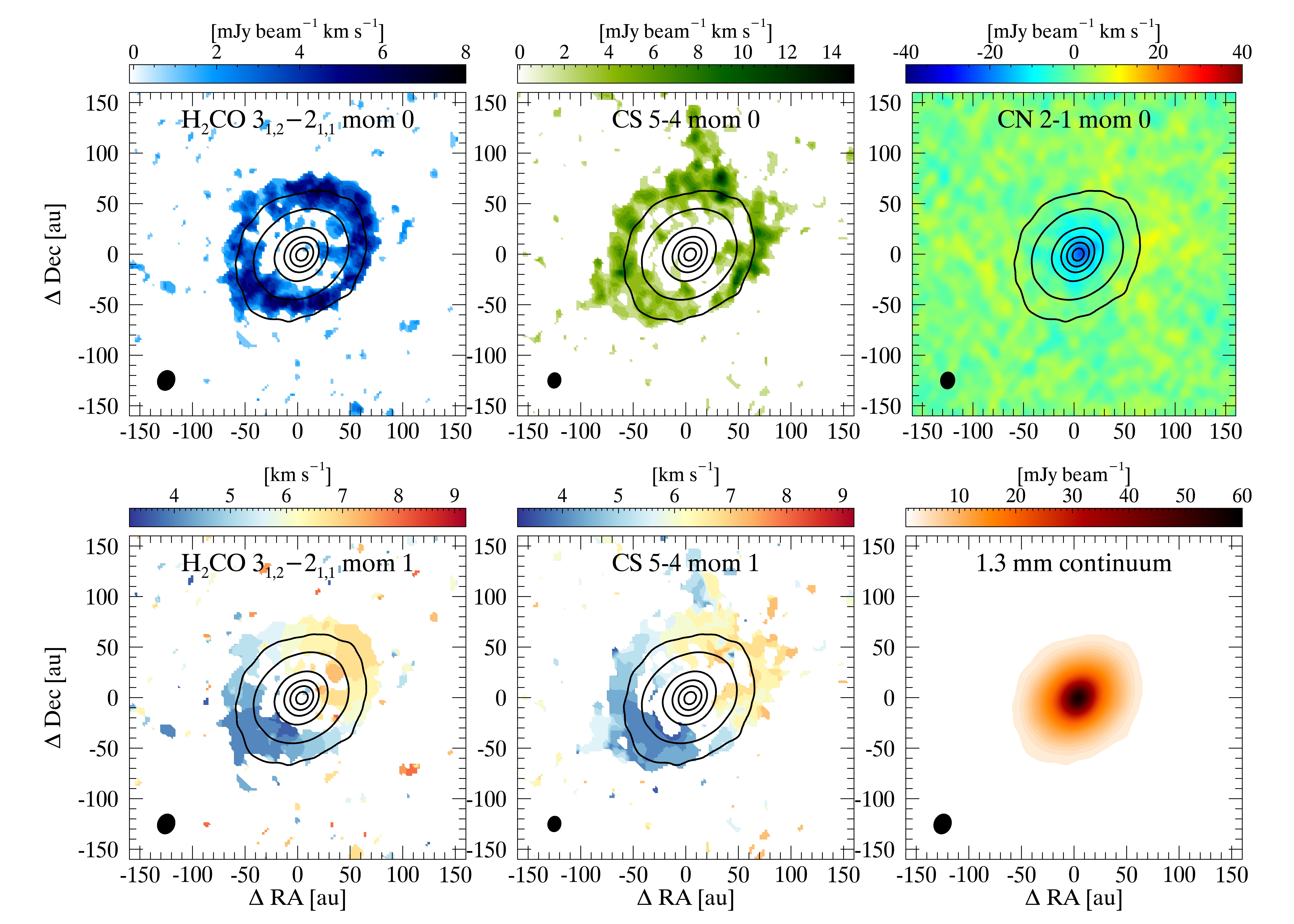}
   \caption{Moment-0   maps (top panels) of H$_2$CO $3_{1,2}-2_{1,1}$ (left), CS $5-4$ (middle), and CN  $2-1$ (right; blending of the three components at 226.87 GHz reported in Table \ref{tab:lines})  towards DG Tau. The corresponding moment-1 maps are shown in the  bottom left and middle panels, while no moment-1 map is shown for CN $2-1$ because of the blending of the three hyperfine components. 
   The bottom right panel shows the map of the continuum emission at 1.3~mm.
   The color scale indicates the line intensity in mJy beam$^{-1} $\kms (moment-0), the velocity V$_{\rm LSR}$ in \kms (moment-1), and the intensity in mJy  beam$^{-1}$ (for the continuum map). The black contours indicate the 1.3~mm dust continuum emission (0.01\% corresponding to the 5$\sigma$ level, 10\%, 30\%, 50\%, 70\%, and 90\% contours). The ellipse in the bottom left corner of each panel shows the ALMA synthesized beam. 
}
              \label{fig:line-moments}
    \end{figure*}

\section{Observations and data products}

\subsection{Observations}

ALMA observations of DG\,Tau were performed during Cycle~4 in August 2017 with baselines ranging from 17\,m to 3.7\,km (project 2016.1.00846.S, PI: L. Podio). The bandpass was calibrated with the quasar J0510+1800, and phase calibration was performed every $\sim$8 minutes using quasar J0438+3004.
The correlator setup consists of 12 high-resolution (0.122\,MHz) spectral windows (SPWs) covering several molecular transitions, among which H$_2$CO $3_{1,2}-2_{1,1}$,  and several hyperfine components of the CN $2-1$ transition,  and one lower resolution spectral window (0.977\,MHz) for the continuum  also covering the  CS $5-4$ line (frequency, $\nu_{0}$, upper level energies, E$_{\rm up}$, and line strengths, S$_{\rm ij} \mu^2$, are listed in Table~\ref{tab:lines}).  
Data reduction was carried out following standard procedures using the ALMA pipeline in CASA 4.7.2. Self-calibration was performed on the source continuum emission by combining a selection of line-free channels and applying the phase-solutions to the continuum-subtracted SPWs.
Continuum images and spectral cubes were produced with "tclean" using an interactive mask on the visible signal until the residuals show no appreciable signal, and a Briggs parameter of 0.5.
The continuum subtraction is  performed by estimating the  continuum  level from the  frequency  range  adjacent  to  the  targeted  lines, that is, from line-free channels.
The flux calibration was performed using the quasars J0238+1636 and J0510+1800, obtaining an accuracy of $\sim 10\%$.
The clean beam, channel width, and r.m.s. over the channel of the resulting line cubes are listed in Table~\ref{tab:lines}.
The continuum image has an r.m.s. of $0.12$ mJy/beam.
\subsection{Data products}

Channel maps of the H$_2$CO $3_{1,2}-2_{1,1}$ and CS $5-4$  emission are shown in Figs. \ref{fig:channel-maps}. For CN $2-1$, the channel maps of the  brightest hyperfine component in our spectral setting ($N=2-1$, $J=5/2-3/2$, $F=7/2-3/2$) is shown in Fig. \ref{fig:cn_channel-maps}.

The moment-0 maps of the H$_2$CO $3_{1,2}-2_{1,1}$ and CS $5-4$ lines were obtained by integrating the line cubes over the velocity channels where emission above the $3\sigma$ level is detected: V$_{\rm LSR} = (+3.24, +9.24)$ \kms, i.e., $\pm 3$ \kms\, with respect to systemic ($V_{\rm sys} = +6.24$ \kms, in agreement with \citealt{podio13,podio19}). A $3\sigma$ clipping was applied. For CN $2-1,$ almost no emission above $3\sigma$ is detected in the channel maps, and therefore the moment-0 map is obtained by integrating on the same velocity interval as for H$_2$CO and CS and no clipping was used. The moment-0 map of CN $2-1$ is due to the blending of the three hyperfine components at 226.87 GHz listed in Table \ref{tab:lines}. The velocity offset between the brightest component ($N=2-1$, $J=5/2-3/2$, $F=7/2-3/2$) and the two fainter ones is  $-1.47$ \kms\, and $+0.78$ \kms. These offsets are smaller than the line broadening due to the disk kinematics. Therefore, the moment-0 map obtained integrating on the disk velocity profile as  defined for the H$_2$CO and CS lines ($V_{\rm sys} \pm 3$ \kms) includes the contributions from the three components. For this reason no moment-1 map was produced for CN $2-1$. The moment-0 and 1 maps are shown in Fig. \ref{fig:line-moments}.

\begin{figure*}
   \centering
\includegraphics[width=15.cm]{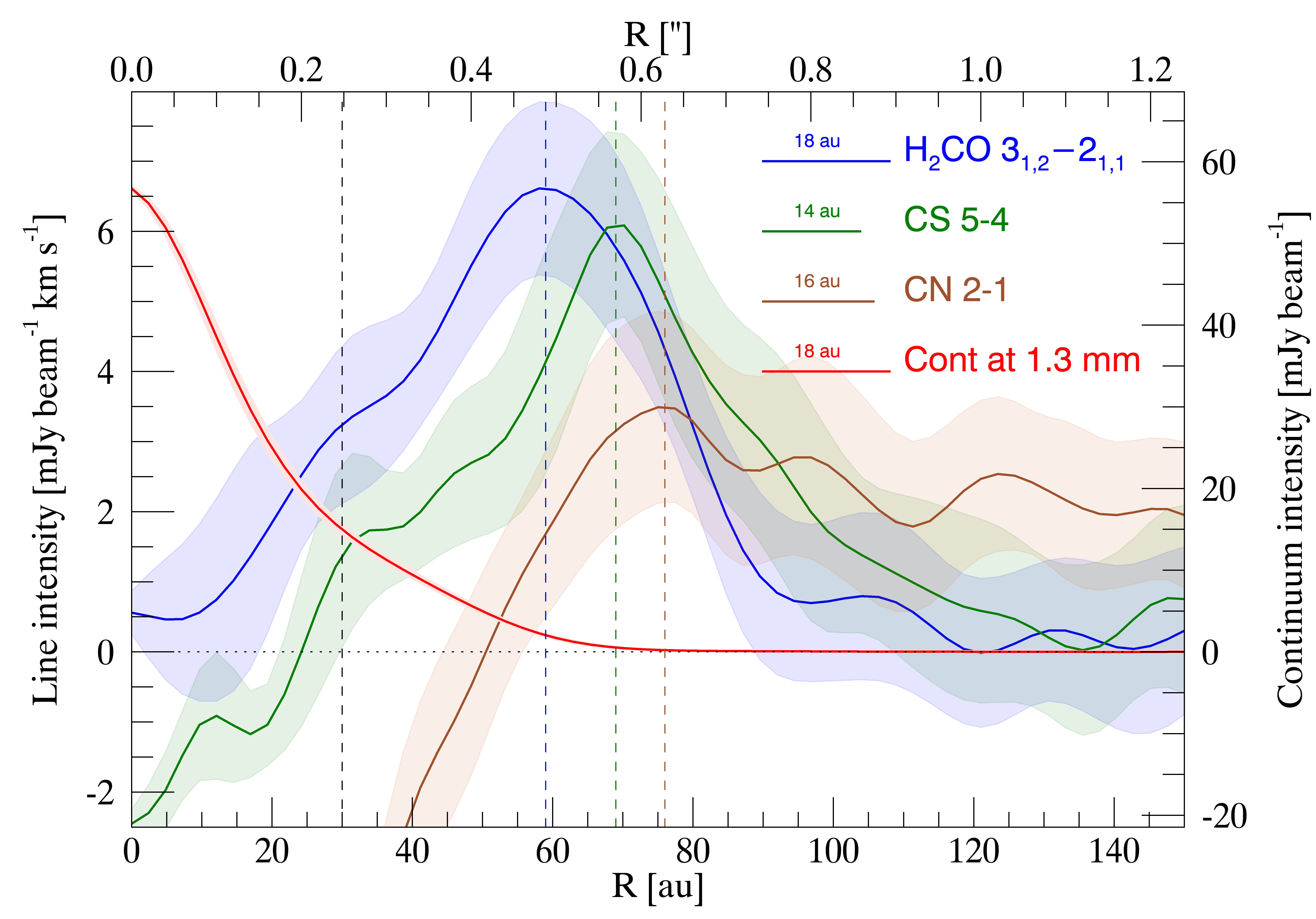}
   \caption{Azimuthally averaged radial intensity profile of H$_2$CO $3_{1,2}-2_{1,1}$ (blue), CS $5-4$ (green), and CN $2-1$ (brown; blending of the three components at 226.87 GHz reported in Table \ref{tab:lines})  in mJy beam$^{-1}$ \kms(left axis), and of the 1.3~mm continuum (red) in mJy beam$^{-1}$( right axis). The radial distance is indicated in au on the bottom axis and arcseconds on the top axis. The angular resolution of the lines/continuum profiles is indicated by the corresponding horizontal lines. The shaded areas indicate the dispersion of the intensity values around the mean along each annulus in the radial direction. The vertical dashed lines indicate the position of the CO snowline (black), and of the peaks of the emission in H$_2$CO $3_{1,2}-2_{1,1}$ (blue), CS $5-4$ (green), and CN $2-1$ (brown).}
              \label{fig:radial-profiles}
    \end{figure*}

The radial intensity profiles of the lines and of the 1.3 mm continuum emission are obtained by azimuthally averaging the unclipped moment-0  maps after deprojecting for
the disk inclination and are shown in Fig. \ref{fig:radial-profiles}.\footnote{With respect to \citet{podio19} we revise the estimate of the disk PA and inclination based on the fit of the self-calibrated continuum map obtained by stacking all the SPWs. We obtain PA$=135\degr$ and i$=35\degr$. However, we note that the extracted azimuthally averaged radial profiles do not change significantly with respect to \citet{podio19}.}
The azimuthal average was performed on the number of resolution elements $N$ enclosing the full angle at separation $r$ ($N=2 \pi r/beam$). The angular resolution of the obtained intensity profiles is set by the mean beam size ($\sqrt{B_{\rm max}\times B_{min}}=0\farcs15$ ($\sim18$ au), $0\farcs11$ ($\sim14$ au), and $0\farcs13$ ($\sim16$ au), for the H$_2$CO, CS, and CN lines, respectively), as indicated in Fig. \ref{fig:radial-profiles}.

\section{Results}

   \begin{figure*}
   \centering
\includegraphics[width=15cm]{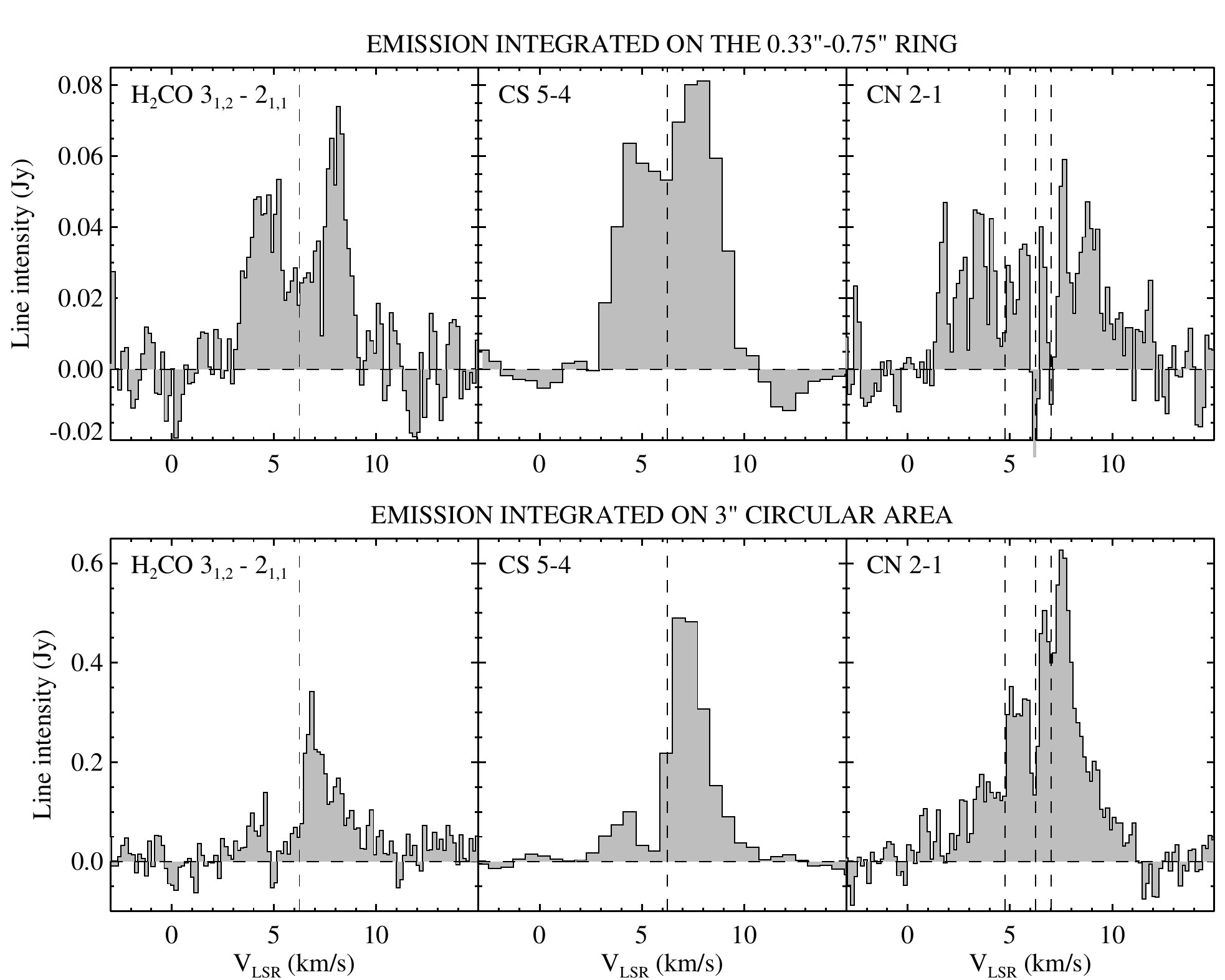}
   \caption{H$_2$CO $3_{1,2}-2_{1,1}$, CS $5-4$, and CN $2-1$  spectra integrated over a $0\farcs33-0\farcs75$ ring area (top panels) and over a $3\arcsec$ circular area (bottom panels). The vertical dashed lines indicate the systemic velocity, V$_{\rm sys} = +6.24$ \kms. For CN $2-1,$ the velocity scale corresponds to the brightest of the three hyperfine components at 226.87 GHz reported in Table \ref{tab:lines} ($N=2-1$, $J=5/2-3/2$, $F=7/2-3/2$), and the position of the two fainter components is indicated by the vertical dashed lines.}
              \label{fig:spec}
    \end{figure*}

\subsection{Spatial distribution of H$_2$CO, CS, and CN emission}
\label{sect:mol}

 We analyze the distributions of the H$_2$CO, CS, and CN emission in the disk of DG Tau from the moment-0 and 1 maps shown in Fig. \ref{fig:line-moments}. The map of H$_2$CO was presented by \citet{podio19} who reported emission from a disk ring located at $\sim 0\farcs33-0\farcs75$ ($40-90$ au) distance from the star, at the edge of the mm continuum emission.

The moment-0 map of Fig.\,\ref{fig:line-moments} shows that the CS emission is roughly co-spatial with that of H$_2$CO. However, the azimuthally averaged radial profiles  of Fig.\,\ref{fig:radial-profiles} reveal that the ring of CS is displaced  outward by $\sim 10$ au; the H$_2$CO emission peaks at $\sim 60$ au and extends out to $\sim 120$ au, while the CS emission peaks at $\sim 70$ au and extends out to $\sim 130$ au. 
The $\sim 10$ au offset between the peaks of the H$_2$CO and CS radial intensity profiles is larger than the uncertainty on the estimate of their position (the position of the intensity peak determined by Gaussian fitting the peak profile is affected by an uncertainty of  approximately one-fifth of the nominal resolution imposed by the beam, that is, $\sim 3.6$ au for H$_2$CO and $\sim 3$ au for CS). 
Therefore, the radial offset between the H$_2$CO and CS emission peaks is real and not due to a resolution effect. 
Being observed along the major axis, this effect cannot be explained by a different vertical origin for the H$_2$CO and CS emissions. At larger radii, the CS emission probes a stream of material extending from the redshifted NW side of the disk out to $\sim 3\arcsec$ towards the north. The NW  side where this stream connects to the disk is the brightest disk region in the continuum emission \citep[see radial cuts in Fig. B.1 of ][]{podio19}, in polarized intensity \citep[Fig. 2 of ][]{bacciotti18}, and in line emission. The asymmetry between the two disk sides in H$_2$CO and CS emission is clearly seen in the moment-0 maps (Fig. \ref{fig:line-moments}), as well as in the line spectra obtained by integrating the line emission over the disk ring as defined by \citet{podio19} ($r \sim 0\farcs33-0\farcs75$ corresponding to $40-90$ au; see Fig.\,\ref{fig:spec}), where the redshifted peak is brighter than the blueshifted one in both lines. The line spectra obtained by integrating the emission out to a radius of $\sim 3\arcsec$, that is, over a region extending well beyond the $40-90$ au disk ring, indicate that this asymmetry is even more pronounced at larger scales. The profiles of CN $2-1$ and H$_2$CO $3_{1,2}-2_{1,1}$ integrated over a 3$"$ region are similar to what is observed with the IRAM-30m by \citet{guilloteau13}. This indicates that outside the disk ring, line emission is likely dominated by the circumstellar envelope, in agreement with what is suggested by \citet{guilloteau13}.\\

The channel maps of the CN $2-1$ emission at 226.87 GHz show negative intensities in the inner disk region, that is, for $r<50$ au, in the channels corresponding to the systemic velocity, for each of the three hyperfine components 
(see Fig. \ref{fig:cn_channel-maps}). For an embedded disk like DG Tau, this is likely due to continuum over-subtraction. Circumstellar material may absorb the disk line emission at $V_{\rm sys}$ as well as the continuum emission at the corresponding frequency. This makes the disk continuum in the frequency channel, which corresponds to the systemic velocity, lower than in the adjacent channels. When continuum subtraction is performed, the continuum level is extracted from the frequency range adjacent to the targeted lines. This  yields a partial over-subtraction at the frequency corresponding to V$_{\rm sys}$. 
Besides the absorption in the inner disk region, no CN $2-1$ emission is detected 
in the outer disk where continuum is fainter or in the channels at blue- and redshifted velocities with respect to $V_{\rm sys}$ which should not be affected by line/continuum absorption by the circumstellar material.
Despite the fact that no clear structures are detected in the channel maps or in the moment-0 map (Figs. \ref{fig:line-moments} and \ref{fig:cn_channel-maps}), the azimuthally averaged radial profile in Fig. \ref{fig:radial-profiles} shows CN emission above the dispersion. The profile indicates that CN emission originates from a more extended region than H$_2$CO and CS, with a peak at $\sim 80$ au and extending out to $\sim 500$ au. However, as the detected CN $2-1$ emission is due to the blending of three hyperfine components, it is difficult to retrieve information on the gas kinematics or to draw conclusions as to  whether the CN emission is associated with the disk or is mostly (or totally) dominated by the residual circumstellar envelope (which also acts to absorb the line/continuum emission from the inner disk). When integrated on the disk ring between $40$ and $90$ au, CN shows a double-peaked profile similar to that of H$_2$CO and CS and consistent with disk emission (see Fig.\,\ref{fig:spec}). However, the spectra integrated on a $3\arcsec$ circular area clearly indicate that the CN emission is strongly affected by extended emission from the residual envelope. \\

\subsection{Dust substructures}
\label{sect:dust}

To highlight the possible presence of dust substructures, we applied an unsharp masking filter to the continuum image at 1.3~mm as well as to that presented by \citet{bacciotti18} at 0.87~mm. Unsharp masking consists in subtracting from the image a blurred version of the same image in order to artificially increase the contrast \citep[see application to disks by e.g.,][]{quanz11,perez16}. Similarly to \citet{garufi16}, here we first divided the original image by the local variance in each pixel and then subtracted the image smoothed by 60 mas. This technique revealed a ring at $\sim40$ au in the continuum maps at 0.87~mm and at 1.3~mm (see Fig.\,\ref{fig:cont+pol}). This structure corresponds to the tentative shoulder identified by \citet{podio19} who extracted the radial intensity profile along the disk major axis and calculated its second derivative (see their Fig. B.1). The location where the second derivative of the intensity profile becomes negative, indicating a local increase of the continuum intensity, is between 38 and 44.5 au, in perfect agreement with the position of the ring revealed by applying unsharp masking. However, the precise morphology of the ring should be interpreted with caution given the arbitrariness of the technique. The barely visible outer ring in the map at 0.87~mm is an artifact of the procedure, and is routinely generated at the outer edge of the detectable signal. Hence, this structure corresponds to the outer edge of the disk emission.

\subsection{Molecular column density}
\label{sect:abu}

In \citet{podio19}, we computed the column density of H$_2$CO in the outer disk ring. 
Here, we followed the same procedure to constrain the column density of CS and CN. We integrated the relative emission over the same area as the H$_2$CO (a circular ring extending from $0\farcs33$ to $0\farcs75$) and converted this integrated flux into column density assuming local thermodynamic equilibrium (LTE) and optically thin emission through the molecular parameters and partition function from the Cologne Database of Molecular Spectroscopy (CDMS, \citealt{muller01}). The assumption of LTE is justified as the gas density in the so-called molecular layer where the lines are thought to originate (see, e.g,  \citealt{oberg17,walsh14}) is high (from $\sim 10^{8}$ to $\sim 10^{12}$ cm$^{-3}$ according to the disk model by \citealt{podio13}), and is well above the critical density of the considered lines (at $20-100$ K, n$_{\rm cr}\sim 7.0-4.6 \times 10^5$ \cmc\, for H$_2$CO $3_{1,2}-2_{1,1}$, $\sim 1.7 \times 10^6 - 9.5 \times 10^5$ \cmc\, for CS $5-4$, and $\sim 2.2 \times 10^6 - 6.4 \times 10^5$ \cmc\, for CN $2-1$, \citealt{shirley15}). Under LTE, the levels are populated according to the Boltzmann distribution and the excitation temperature is equal to the gas kinetic temperature $T_{\rm ex} = T_{\rm K}$. The temperature of the emitting layer has to be assumed as we have only one line per species. Recent multi-line studies of CS and H$_2$CO emission in a few protoplanetary disks indicate that the lines originate from a disk layer at or above the CO freeze-out temperatures \citep{legal19,pegues20,teague18}. Therefore, we assume a lower range of temperatures with respect to what was previously assumed by \citet{podio19} and compute the column density of H$_2$CO, CS, and CN for $T_{\rm ex} = T_{\rm K} = 20-100$ K.
This procedure yielded ring- and disk-height averaged column densities of $1.8 - 5.5 \times 10^{13}$ cm$^{-2}$ for o-H$_2$CO, which translates into a total column density  of $2.4-8.6 \times 10^{13}$ cm$^{-2}$ for  H$_2$CO based on the o/p ratio estimated by \citet{guzman18a}, and of $1.7 - 2.5 \times 10^{13}$ cm$^{-2}$ for CS. For CN the integrated fluxes are obtained from the moment-0 maps of the CN $2-1$ components at 226.66 and 226.87 GHz. From the brightest lines at 226.87 GHz a column density of $1.9 - 4.7\times 10^{13}$ cm$^{-2}$ is estimated, which is consistent with the upper limit derived from the fainter components at 226.66 GHz. The derived column density values are in agreement with the range of values found for other disks (\citealt{carney19,pegues20} for H$_2$CO, \citealt{legal19,teague18} for CS, and \citealt{hilyblant17} for CN, \citealt{garufi20b,podio20a,vanthoff20} for the three molecules). The integrated line intensities and the derived column densities are summarized in Table \ref{tab:lines}.   

As shown by recent observational studies \citep{legal19,pegues20,teague18}, CS and H$_2$CO emission may be optically thick in protoplanetary disks. In order to check whether or not and to what extent line opacity may affect our column density estimates, which are derived assuming optically thin emission, we converted the ring-integrated line spectra shown in Fig. \ref{fig:spec} in brightness temperature (see Fig. \ref{fig:spec-K} in Appendix \ref{app:spec-K}). The line brightness temperature, $T_{\rm B}$, is well below the gas temperature for optically thin lines, while it is $\sim (0.6-0.8) \times T_{\rm K}$ for thermalized lines.
We find that the intensity peak of the H$_2$CO $3_{1,2}-2_{1,1}$ and CS $5-4$ integrated spectra is T$_{\rm B} \sim 1.5$ K. This value is well below the typical gas temperatures of the H$_2$CO and CS emitting disk layer ($T_{\rm ex} \sim T_{\rm K} \sim 11-37$ K for H$_2$CO \citep{pegues20}, and $\sim 20-35$ K for CS \citep{legal19,teague18}). Thus, we conclude that the observed lines are likely optically thin.
\citet{legal19} and \citet{pegues20} find that H$_2$CO and CS emission is mildly optically thick in the protoplanetary disks MWC 480 and LKCa 15, and that  H$_2$CO  is also  mildly optically thick in    DM Tau and J1604-2130. These latter authors infer $\tau \sim 0.07-0.4$ for CS $5-4$ and $\tau \sim 0.14-0.85$ for the H$_2$CO lines with the lower upper level energies, that is, $E_{\rm up} \sim 20-34$ K, and therefore the column densities derived using the optically thin approximation are underestimated by a factor of about two. As we cannot exclude that the detected H$_2$CO and CS emission lines are mildly optically thick in the disk of DG Tau, our estimates of the column densities may also be underestimated by a factor of a few.

\section{Discussion}

\subsection{The origin of ring-like emission}

The H$_2$CO, CS, and CN moment-0 maps and radial intensity profiles show that the bulk of the line emission originates from an outer disk ring while poor emission (or even negative fluxes in the case of CN and CS) is detected in the inner disk region, that is, at radii $\le 50$ au.  
As discussed in \citet{podio19}, the depression or lack of emission in the inner disk region may be due to: (i) lower molecular column density in the inner disk region; (ii) optically thick dust suppressing line emission; and (iii) absorption of the dust continuum emission by optically thick line emission from the disk itself and/or from circumstellar material, which would then lead to an over-subtraction at the line frequency when removing the continuum evaluated in the frequency range adjacent to the line. As discussed in Sect. \ref{sect:mol}, the latter may 
be the cause of the negative values seen in the inner disk region in the channel maps, the moment-0 map, the radial profile, and the integrated spectra of CN $2-1$.
Negative fluxes in the inner 20 au are also seen in the radial intensity profile of CS $5-4$, while no negative values are seen in the H$_2$CO profile.  
Following \citet{podio19} we exclude that the depression in the H$_2$CO and CS emission is due to dust opacity for $r>20$ au. This is based on previous modeling of the continuum emission that shows that the disk is optically thick in the inner $10$ au. Further out, the optical depth sharply decreases to values lower than $0.5$ at $\sim20$ au \citep{isella10}. 
Furthermore, if the inner depression was due to optically thick dust, the observed molecular rings would be centered around the continuum peak. Instead, \citet{podio19} noted that the H$_2$CO ring is displaced along the major axis with respect to the continuum and the same effect is visible in the CS map presented in this work (see Fig.\,\ref{fig:line-moments}).
Finally, in the case of optically thick dust, all the lines should be equally suppressed at the same radii whereas the inner profile of our lines is different (see Fig.\,\ref{fig:radial-profiles}), with the CO isotopologs, $^{13}$CO and C$^{18}$O, which show a smaller hole in their distribution compared to H$_2$CO (of $\sim 25$ au, \citealt{gudel18}). This behavior is  the opposite of what was observed for DG Tau B by \citet{garufi20b}.

In conclusion, while we cannot exclude the presence of H$_2$CO and CS emission in the inner $25$ au disk region, which would remain undetected due to optically thick dust and/or continuum over-subtraction due to optically thick line emission, we confirm the presence of an outer ring of enhanced molecular emission at the edge of the mm continuum. An inner hole or dip in the distribution of H$_2$CO was also detected in other disks, namely TW Hya \citep{oberg17}, HD 163296 \citep{carney17}, V4046 Sgr \citep{kastner18}, and 5 out of 13 disks associated with H$_2$CO emission in the survey of (\citealt{pegues20}; two of which are HD 163296 and V4046 Sgr). However,  for all disks except IM Lup and HD 163296, optically thick dust is likely not the cause of the observed lack of H$_2$CO in the inner disk region, similarly to what we conclude for the disk of DG Tau. Moreover, a peak of the H$_2$CO emission at the edge of the millimeter continuum is found in both TW Hya and HD 163296, as well as in 4 of the 13 disks with detected H$_2$CO in the survey by \citet{pegues20}. Concerning CS, the studies by \citet{legal19} and \citet{loomis20} show centrally peaked CS emission in all but one out of five disks and no emission bump at the edge of the continuum. An emission peak at the edge of the dusty disk has also been reported for other molecules, such as  DCO$^+$ and CO isotopologs \citep[see, e.g., ][]{favre19,huang16,oberg17}.  

Concerning CN, as discussed in Sect. \ref{sect:mol},  almost no emission is detected  in the channel maps and in the moment-0 map, but CN emission is clearly visible in the azimuthally averaged radial profile and in the integrated spectra (Figs. \ref{fig:radial-profiles} and \ref{fig:spec}). It is not clear if the emission originates at least partially from an outer disk region or if it is mostly from the envelope. In the second case, the negative fluxes in the inner disk region could be due to absorption of the disk continuum emission by the surrounding extended envelope, which would mask any disk emission.

\subsection{The interplay of gas and dust}

\begin{figure*}
   \centering
\includegraphics[width=15cm]{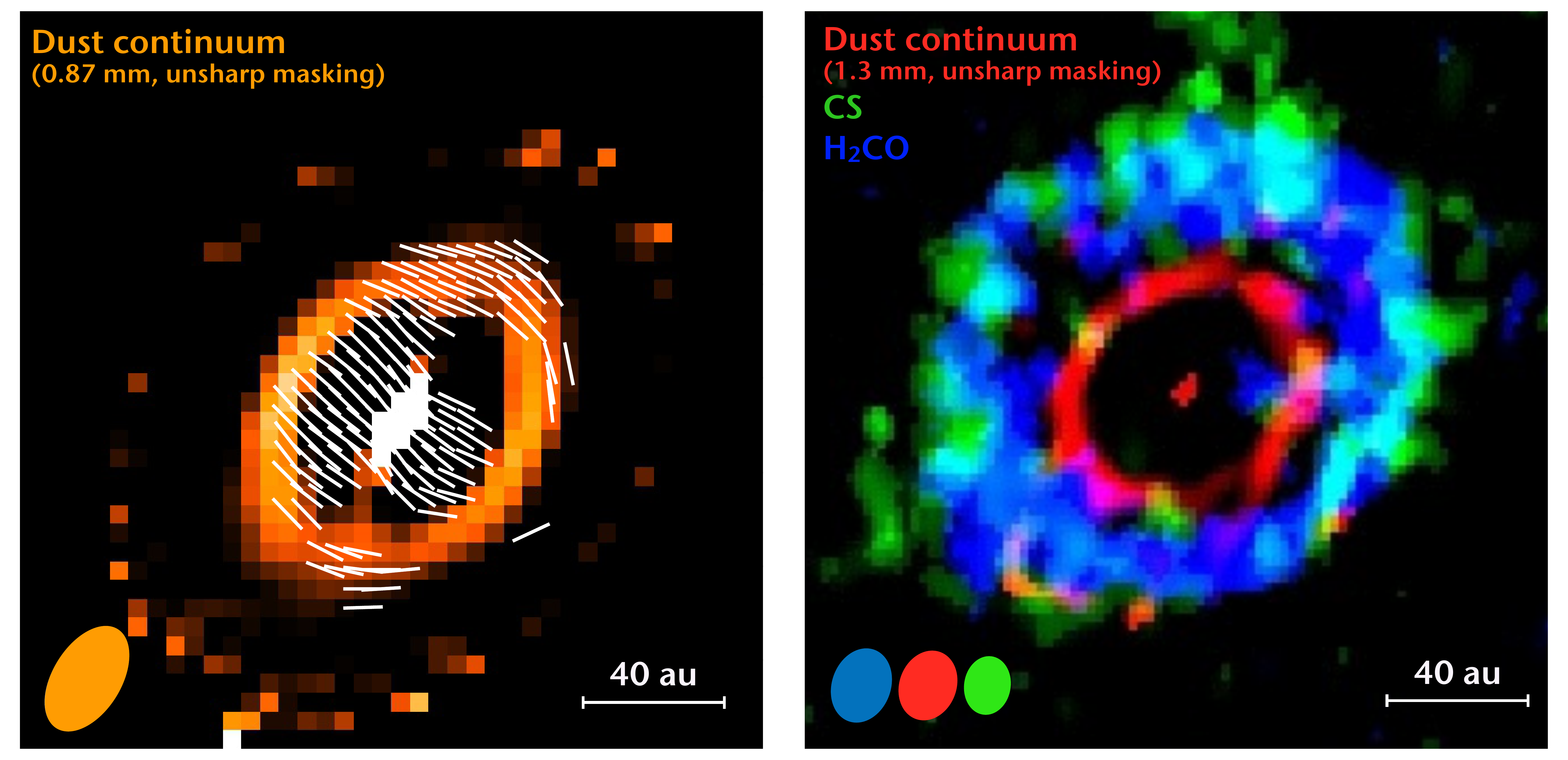}
   \caption{{\it Left panel}: Continuum map at 0.87~mm  after unsharp masking compared with the orientation of the linear polarization vectors at 0.87~mm \citep{bacciotti18}.  The orange ellipse in the bottom left corner indicates the synthesized beam of the 0.87~mm continuum. {\it Right panel}: RGB image showing the continuum at 1.3~mm after unsharp masking (in red),  H$_2$CO $3_{1,2}-2_{1,1}$ (in blue), and CS $5-4$ (in green). The blue, red, and green ellipses in the bottom left corner indicate the synthesized beam for H$_2$CO, continuum, and CS, respectively.}
              \label{fig:cont+pol}
    \end{figure*}

To quantify the degree of dust accumulation occurring at the location of the dust ring revealed by unsharp masking (see Fig.\,\ref{fig:cont+pol}), we integrated the continuum flux encompassed in the original map at the ring location as well as the total flux from the disk. A Gaussian fit to the observed emission yields a disk major axis of $0\farcs45$ and a disk inclination $i$ of 35$\degr$. The flux integrated over this region is 0.29 Jy whereas the ring flux amounts to 0.04 Jy. It is therefore clear that this substructure is morphologically different from the bright structures recurrently imaged by ALMA \citep[see e.g.,][]{andrews18}. Indeed, the dust ring in the disk of DG Tau only shows a $\sim10\%$ flux enhancement with respect to the contiguous disk regions and this converts into a comparable, and thus marginal, amount of dust accumulation (assuming that the enhanced emission is not due to changes in the dust opacity and/or temperature in the ring). The dust mass of disk and ring can be obtained from the aforementioned fluxes in the assumption of optically thin emission through:
\begin{equation}
M_{\rm dust}=\frac{F_{\rm mm}d^2}{\kappa_{\lambda}B_{\lambda}(T_{\rm dust})}
,\end{equation} 
where the dust opacity $\kappa_{\lambda}$ at 1.3 mm is assumed to be a global 2.3 $\rm cm^2\,g^{-1}$ \citep[from][]{beckwith90} and the Planck function $B_{\lambda}(T_{\rm dust})$ is calculated from an overall dust temperature $T_{\rm dust}$ of 20 K. Under these assumptions, the total dust mass is $\sim 120$ M$_{\oplus}$ and the ring dust mass is $\approx16\,{\rm M_{\oplus}}$.

The dusty ring visible after unsharp masking shows spatial analogies with both the linear polarization map at 0.87 mm by \citet{bacciotti18} and the molecular emission presented in this work, as shown in  Fig.\,\ref{fig:cont+pol}. The orientation of the polarization vectors changes from parallel to the disk minor axis to azimuthal at the radial location of the ring, that is, at $\sim 40$ au. This confirms the suggestion made by \citet{bacciotti18} that sharp changes in the orientation of the polarization pattern may betray the presence of yet unseen substructures in the dust and gas distribution. Moreover, both the CS and H$_2$CO rings of emission of Fig.\,\ref{fig:line-moments} lie at the outer edge of the continuum substructure. We also note that the dusty ring is located beyond the CO snowline at $R_{\rm CO} \sim 30$ au, as computed by the thermo-chemical disk model ProDiMo \citep{podio13}. This suggests that outside the CO snowline there could be a change in the dust properties (e.g., dust grain size and opacity) which would lead to (i) an enhancement of the dust continuum emission; (ii) a change of the orientation of the polarization due to dust grain self-scattering \citep{bacciotti18}; and (iii) an enhancement of H$_2$CO and CS emission.

\subsection{The chemical origin of detected molecules}


The observed H$_2$CO $3_{1,2}-2_{1,1}$ and CS $5-4$ emission originate from  roughly the same radial region of the disk, that is, from a ring located between $\sim 40$ au and $\sim 130$ au and with a peak close to the edge of the distribution of the mm dust grains, with the CS emission displaced by $\sim 10$ au towards the outer disk. 
The two transitions have very similar upper level energy and critical density ($E_{\rm up}=33$ K, and 35 K, and $n_{\rm cr} \sim 7-5 \times 10^5$ \cmc, and $17-9.5 \times 10^5$ \cmc\, at $20-100$ K, for the H$_2$CO and the CS line, respectively). Therefore, they are excited in similar conditions and it is reasonable to assume that they 
also arise from the same vertical region of the disk, in agreement with disk modeling by \citet{fedele20} and with the ALMA images of the edge-on disk of IRAS 04302+2247 \citep{podio20a}.

H$_2$CO and CS could also be linked from a chemical point of view. Following \citet{legal19} the main formation routes of CS are either (i)  rapid ion--neutral reactions between S$^+$ and small hydrocarbons (such as CH$_x$ and C$_y$H, with $x=1-4$ and $y=2-3$), which produce carbonated S-ions, including HCS$^+$, CS$^+$, HC$_3$S$^+$, and C$_2$S$^+$, which subsequently recombine with electrons to form neutral S-bearing species;  or (ii) neutral–neutral reactions between S and small hydrocarbons (at deeper disk layers). The main formation route of H$_2$CO in gas phase is through the reaction CH$_3$ + O, which is efficient in the warm inner region and molecular layers of the disk where atomic oxygen is produced by photodissociation of gas-phase CO \citep[e.g., ][]{loomis15,willacy09}. The latter also makes available C for the formation of small hydrocarbons which boost the formation of both H$_2$CO and CS. Finally, both species are easily destroyed in the disk atmosphere due to photo-dissociation and  reactions with protons and protonated ions (i.e., with H$^+$, H$_3^+$, and HCO$^+$), and will freeze-out onto dust grains in the disk midplane. The freeze-out will occur when the dust temperature falls below their freeze-out temperature ($T_{\rm d}$) which is $\sim 65$ K for CS and $\sim 90$ K for H$_2$CO as estimated from the binding energy by \citet{wakelam17} ($E_{\rm b}$=3200 K and 4500 K, respectively)\footnote{The binding energies (BE) of CS and H$_2$CO by \citet{wakelam17} are in agreement with the range estimated by \citet{ferrero20} and \citet{codella20}, while} lower binding energies and freeze-out temperatures are found by \citet{penteado17}. 
Therefore, if H$_2$CO and CS are primarily produced in gas-phase in the disk molecular layer, their abundance would be strongly linked to the presence of small hydrocarbons, such as CH$_3$.
On the other hand, there could be a second reservoir of H$_2$CO and CS due to the release from the dust-grain mantles. For both H$_2$CO and CS, the peak of emission at the edge of the millimeter continuum could be due to several mechanisms, as discussed by \citet{carney17}, \citet{oberg17}, \citet{pegues20}, and \citet{podio19}: (i) enhanced photodesorption from dust grains; (ii) enhanced photodissociation of CO, which produces atomic O and small hydrocarbons for the formation of H$_2$CO, as well as CS; or (iii) temperature inversion in the outer disk region with fewer solids. 
 
With respect to H$_2$CO and CS, CN emission originates from a larger ring located outside the dusty disk, and extending  from $\sim 80$ au to $\sim 500$ au. CN ring-like emission was detected in a few other disks, namely in TW Hya \citep{hilyblant17}, and the Sz 71 and Sz 68 disks in Lupus \citep{vanterwisga19}. 
As discussed by \citet{cazzoletti18}, the main reactions that form CN start from H$_2^*$, that is, H$_2$ molecules that are pumped into excited vibrational states by far-ultraviolet (FUV) radiation. The reaction N + H$_2^*$ $\to$ NH +H is followed by C$^+$ + NH $\to$ CN$^+$ +H, and CN$^+$ finally goes to CN through a charge transfer with H, or following dissociative recombination of the intermediates HCN$^+$ and HCNH$^+$. As the abundance of H$_2^*$ strongly depends on the FUV radiation, CN could be a good tracer of the outer disk surface layers which are more strongly exposed to intense FUV irradiation, as predicted by disk thermo-chemical models \citep{cazzoletti18,fedele20}, and in agreement with the CN ring-like morphology detected in the disks of TW Hya, Sz 71, and Sz 68 \citep{hilyblant17,vanterwisga19}. However, a recent study by \citet{arulanantham20} shows that the intensity of the CN lines detected at mm wavelengths are anticorrelated with the FUV continuum measured from the HST spectra. This may be due to the fact that although FUV irradiation promotes CN formation due to increased production of H$_2^*$ and atomic N,  it also increases CN destruction. In light of this, in the case of DG Tau, the detected CN emission is likely to originate mostly from the circumstellar envelope, and only in small part from the UV irradiated outer disk layers.


\section{Conclusions}

ALMA observations at $0\farcs15$ resolution of the disk of DG Tau show that H$_2$CO and CS emission originates from a disk ring located at the edge of the 1.3~mm dust continuum (R$_{\rm dust} \sim 66$ au), with the peak of CS emission being found at $\sim 70$ au, that is, slightly outside the H$_2$CO peak at $\sim 60$ au. The fact that H$_2$CO and CS emission are roughly co-spatial suggests that CS and H$_2$CO molecules are chemically linked, as both of them may be formed in gas-phase from simple radicals, and/or desorbed from dust grain mantles. Moreover, as the considered CS and H$_2$CO lines have similar excitation conditions it is likely that they originate from the same disk region both radially and vertically. The CN emission emerges outside the 1.3~mm dust emission ($\sim 80$ au) and extends out to $\sim 500$ au. The anti-correlation between H$_2$CO, CS, and CN line emission and the dust continuum could be due to (i) dust opacity and/or over-subtraction of the continuum (due to optically thick circumstellar gas) which screen line emission from the inner disk; and/or (ii) the fact that the outer disk at the edge of/outside the mm dust is more exposed to UV radiation which may enhance the abundance of simple radicals, atomic O, and H$_2^{*}$, thereby enhancing gas-phase formation of CS, H$_2$CO, and CN, as well as  photodesorption of molecules from grains. Some features of the observed molecular emission, such as the co-spatiality of H$_2$CO and CS emission, CN emission extending out to larger radii, is in agreement with the predictions of thermo-chemical models \citep{cazzoletti18,fedele20}.
After unsharp masking, the continuum at 0.87~mm and 1.3~mm shows a ring of enhanced dust emission  located at $\sim 40$ au, that is,  outside the CO snowline ($\sim 30$ au). Interestingly, the peak of molecular emission is just outside this ring of enhanced dust emission where we also observe a change in the  orientation of the linear polarization at 0.87~mm. This suggests a change in the dust properties outside the CO snowline and a link between the observed molecular emission and the dust properties.

\begin{acknowledgements}
     This paper uses ALMA data from project 2016.1.00846.S (PI: L. Podio). ALMA is a partnership of ESO (representing its member states), NSF (USA) and NINS (Japan), together with NRC (Canada), MOST and ASIAA (Taiwan), and KASI (Republic of Korea), in cooperation with the Republic of Chile. The Joint ALMA Observatory is operated by ESO, AUI/NRAO and NAOJ. This work was partly supported by PRIN-INAF/2016 GENESIS-SKA and by the Italian Ministero dell'Istruzione, Universit\`a e Ricerca, through the grants Progetti Premiali 2012/iALMA (CUP-C52I13000140001), 2017/FRONTIERA (CUP-C61I15000000001), SIR-(RBSI14ZRHR), and by the European MARIE SKLODOWSKA-CURIE ACTIONS under the European Union’s Horizon 2020 research and innovation programme, for the Project "Astro-Chemistry Origins" (ACO), Grant No 811312.
     CC, CF, and EB acknowledge funding from the European Research Council (ERC) under the European Unions Horizon 2020 research and innovation programme, for the Project The Dawn of Organic Chemistry (DOC), grant agreement No 741002. CF also acknowledges financial support from the French National Research Agency in the framework of the Investissements d’Avenir program (ANR-15-IDEX-02), through the funding of the "Origin of Life" project of the Univ. Grenoble-Alpes.
\end{acknowledgements}

   \bibliographystyle{aa} 
   \bibliography{mybibtex.bib} 

\begin{appendix}

\section{Channel maps}

\begin{figure*}
   \centering
\includegraphics[width=17cm]{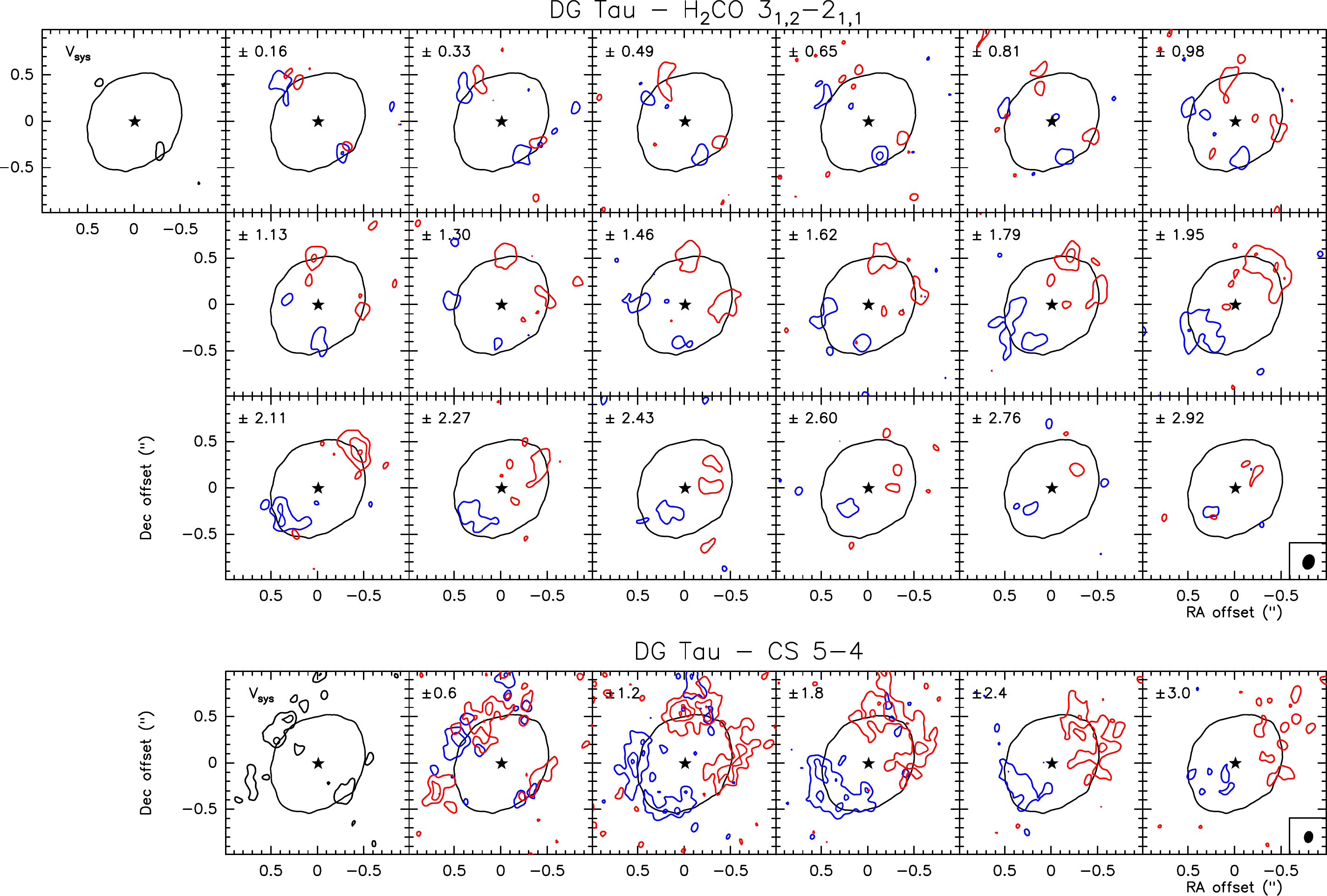}
   \caption{Channel maps of H$_2$CO $3_{1,2}-2_{1,1}$ and CS $5-4$ emission towards DG Tau. The blue and red contours show the emission at symmetric blue- and redshifted velocities with respect to systemic velocity (V$_{\rm sys} = +6.24$ \kms) up to $V-V_{\rm sys} = \pm 3$ \kms, as labeled in the upper left corner of each channel box. The H$_2$CO $3_{1,2}-2_{1,1}$ and CS $5-4$ lines are observed in the  narrow (0.162 \kms\, per channel) and broad (0.6 \kms\, per channel) spectral window, respectively. First contour and step are 3$\sigma$. The black star and contour indicates the peak and the 5$\sigma$ level of the 1.3 mm continuum. The ellipse in the bottom right corner of the last channel shows the ALMA synthesized beam.}
              \label{fig:channel-maps}
    \end{figure*}

\begin{figure*}
   \centering
\includegraphics[width=17cm]{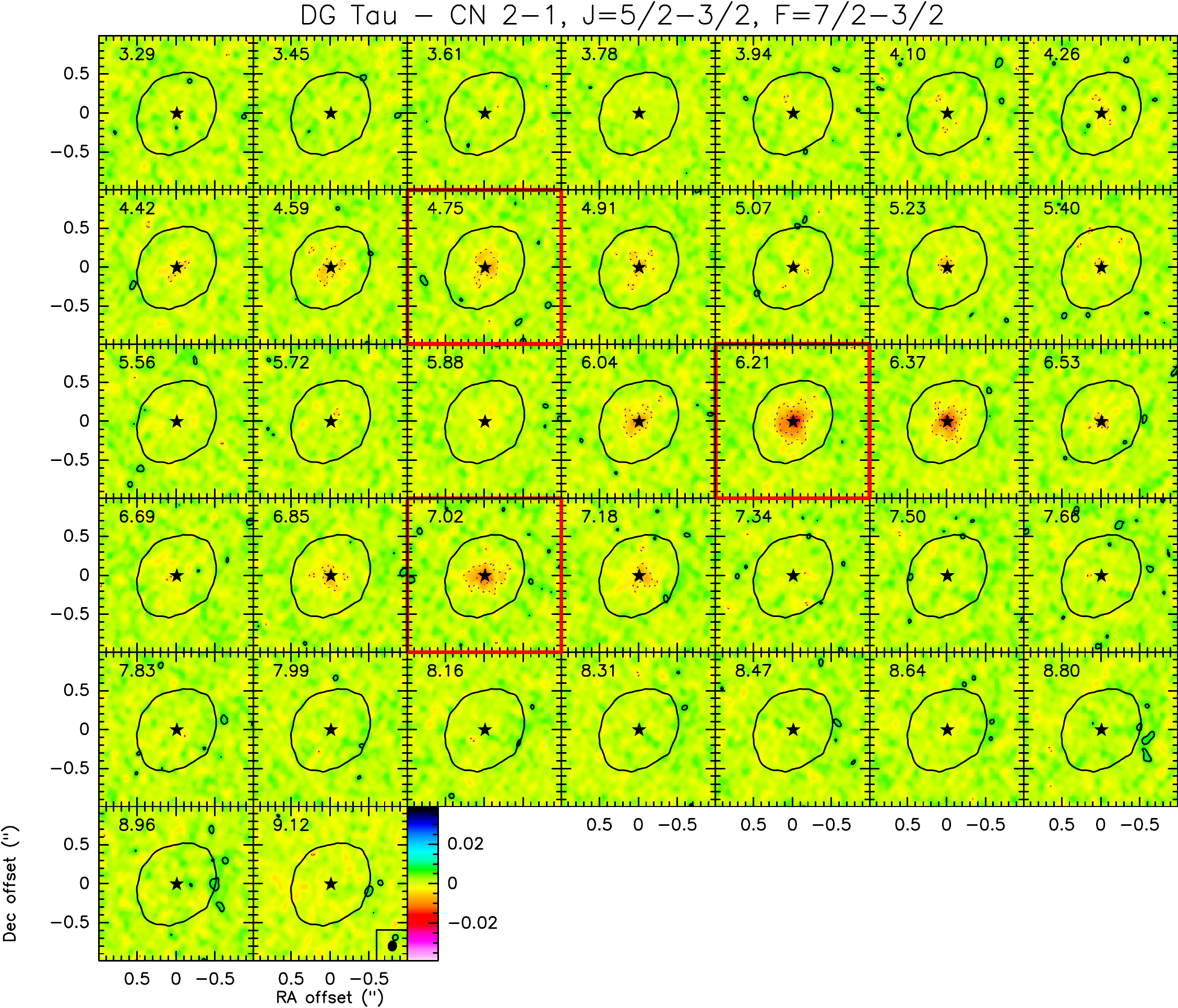}
   \caption{Channel maps of CN $2-1$ emission towards DG Tau. The channel velocity, $V_{\rm LSR}$, is relative to the brightest CN $2-1$, J=5/2-3/2, F=7/2-3/2 hyperfine component and is labeled in the upper left corner of each channel box. The other two hyperfine components are offset by $-1.47$ \kms\, and $+0.78$ \kms\, with respect to the brightest one (see Table \ref{tab:lines}). The channels corresponding to the systemic velocity for the three hyperfine components (i.e. $6.21$ \kms, $4.75$ \kms, and $7.02$ \kms) are highlighted by a thick red box. First contour and step are 3$\sigma$, negative intensities are shown by dashed contours. The black star and contour indicates the peak and the 5$\sigma$ level of the 1.3 mm continuum. The ellipse in the bottom right corner of the last channel shows the ALMA synthesized beam.}
              \label{fig:cn_channel-maps}
    \end{figure*}

The channel maps of the  H$_2$CO $3_{1,2}-2_{1,1}$, CS $5-4$, and CN $2-1$ emission towards DG Tau are presented in Fig. \ref{fig:channel-maps} and \ref{fig:cn_channel-maps}.

\section{Integrated spectra in brightness temperature}
\label{app:spec-K}

\begin{figure*}
   \centering
\includegraphics[width=15cm]{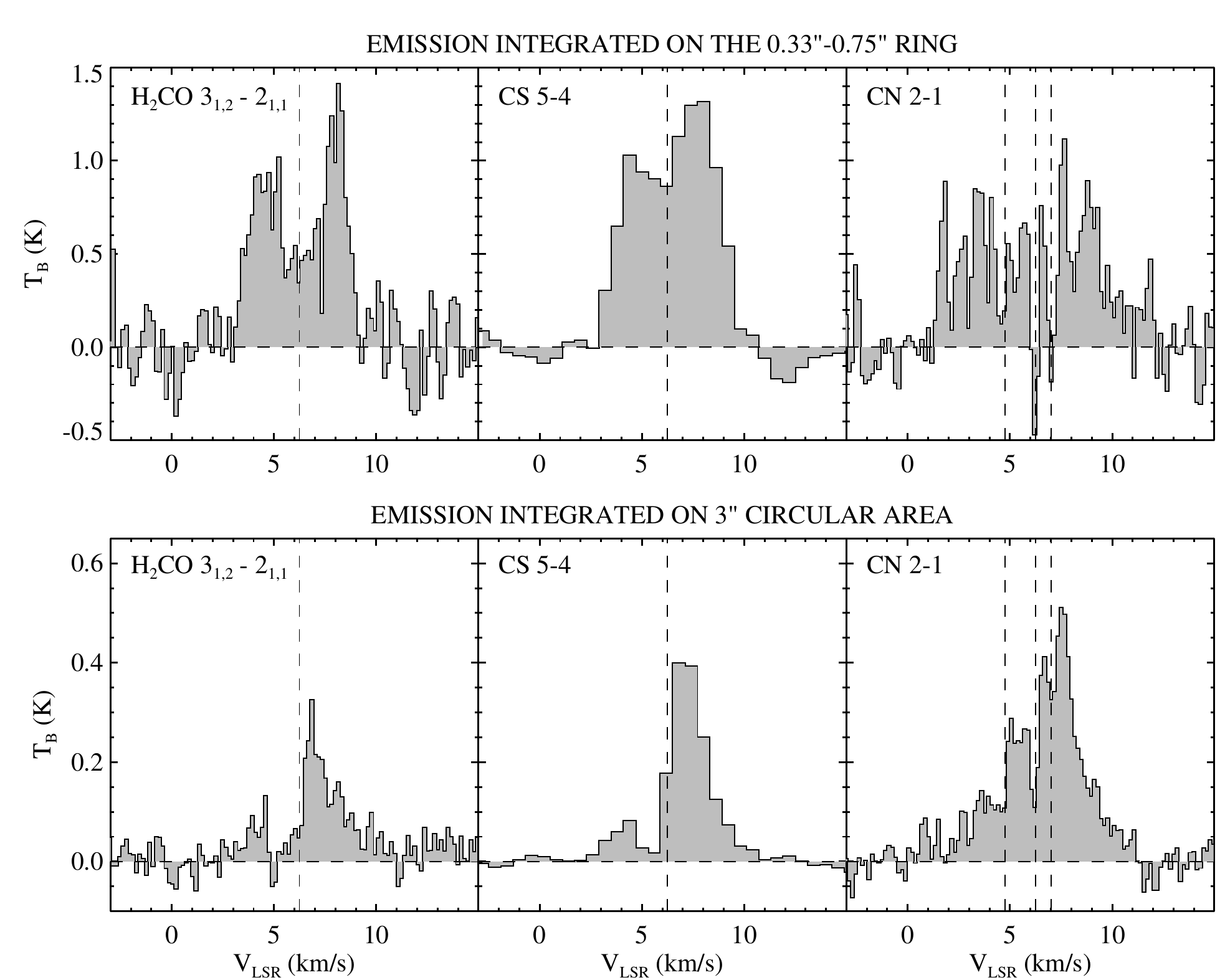}
   \caption{H$_2$CO $3_{1,2}-2_{1,1}$, CS $5-4$, and CN $2-1$  spectra integrated over a $0\farcs33-0\farcs75$ ring area (top panels) and over a $3\arcsec$ circular area (bottom panels), in brightness temperature $T_{\rm B}$. The vertical dashed lines indicate the systemic velocity, V$_{\rm sys} = +6.24$ \kms. For CN $2-1$ the velocity scale corresponds to the brightest of the three hyperfine components at 226.87 GHz reported in Table \ref{tab:lines} ($N=2-1$, $J=5/2-3/2$, $F=7/2-3/2$), and the position of the two fainter components is indicated by the vertical dashed lines.
   }
              \label{fig:spec-K}
    \end{figure*}

The line spectra shown in Fig. \ref{fig:spec} obtained by integrating the line cube over a disk ring extending from $0\farcs33$ to $0\farcs75$ and over a circular area of $3\arcsec$ are converted in brightness temperature $T_{\rm B}$ as shown in Fig.~\ref{fig:spec-K}.

\end{appendix}

\end{document}